\documentclass{article}
\usepackage{aas2pp4,tighten}

\def\alwaysmath#1{\ifmmode{#1}\else{$#1$}\fi}
\def\msun{\alwaysmath{{M}_{\odot}}}
\def\teff{\alwaysmath{T_{\rm eff}}}
\def\logg{\alwaysmath{\log g}}
\def\kelvin{{\rm \,K}}

\newcommand\hst{{\it HST}}
\def\lhb{\alwaysmath{\ell_{\scriptscriptstyle\rm HB}}}

\def\msun{\alwaysmath{\,{M}_{\odot}}}
\newcommand\ngc[1]{NGC\,#1}
\newcommand\etal{et~al.}
\def\feh{{\rm [Fe/H]}}

\def\SP{$2^{\rm nd}$P}
\def\FP{$1^{\rm st}$P}
\def\btsp{BT-$2^{\rm nd}$P}
\def\process{process-${\cal{BT}}$}
\def\manque{AGB-manqu\'e}
\newcommand\etaml{\ensuremath{\eta_{\scriptscriptstyle\rm ML}}}
\slugcomment{Submitted to ApJ}

\lefthead{Ferraro et al.}
\righthead{Multimodal Distributions}

\begin{document}

\title{Multimodal Distributions along the Horizontal Branch
\footnote{Based on
observations with the NASA/ESA {\it Hubble Space Telescope}, obtained at
the Space Telescope Science Institute, which is operated by AURA, Inc.,
under NASA contract NAS5-26555}}

\author{Francesco R. Ferraro\altaffilmark{2},  
Barbara Paltrinieri\altaffilmark{2},  
Flavio Fusi Pecci\altaffilmark{2,3}}
\author{Ben Dorman\altaffilmark{4,5}
and Robert T. Rood\altaffilmark{4}} 

\altaffiltext{2}{Osservatorio Astronomico di Bologna, via Zamboni 33, I-40126
Bologna, ITALY}
\altaffiltext{3}{Stazione Astronomica di Cagliari, 09012 Capoterra, ITALY}
\altaffiltext{4}{Astronomy Dept, University of Virginia,
	P.O.Box 3818, Charlottesville, VA 22903-0818}
\altaffiltext{5}{Laboratory for Astronomy \& Solar Physics, 
Code 681, NASA/GSFC,	Greenbelt MD 20771}

\begin{abstract}

We report on HST/WFPC2 $U,~V$ and far-ultraviolet observations of two
Galactic Globular Clusters (GGCs), \ngc{6205} $=$ M13 and \ngc{6093}
$=$ M80.  Both of these clusters have horizontal-branch (HB) tails
that extend to the helium-burning main sequence, with the hottest
stars reaching theoretical effective temperatures above 35,000\kelvin.
In both clusters, groups of stars are found to be separated by narrow
gaps along the blue HB sequence. These gaps appear at similar
locations in the color-magnitude diagrams of the two clusters. While
stochastic effects may give rise to variations in the color
distribution along the HB, the coincidence of gaps in different
clusters effectively rules this out as the primary cause.  The
comparison among the clusters strongly suggests that there are
separate physical processes operating during the earlier red-giant
phase of evolution to produce mass loss.

\end{abstract}


\section{Introduction}

A growing number of Galactic Globular Clusters (GGCs) have been found
to show discontinuous stellar distributions along the
horizontal-branch (HB).  Some of them are clearly bimodal with a HB
clump of red stars and a tail of blue, hot stars populating the blue
side of the RR Lyrae strip (\ngc{2808}---\cite{ferraro2808};
\ngc{1851}---\cite{wal}; \ngc{6229}---\cite{bori}) and more recently
\ngc{362} (\cite{ngc362}), \ngc{6388} and \ngc{6441}  
(\cite{rich97}).  In other clusters the stellar distribution in the HB
blue tails (BTs) is interrupted by underpopulated regions or ``gaps''
(see for example \ngc{6752}---\cite{buo86}; M15---\cite{bcf85}).
\ngc{2808} (\cite{sos97}) shows both phenomena: it has both a red
clump and an extended blue sequence with three groups separated by
narrow gaps.

The maximum possible extent of the HB tail is $\sim 4.5\,$mag where it
reaches the He-burning main sequence. This limit is
reached in a number of clusters, including \ngc{6752}, \ngc{2808},
$\omega $~Cen and the subjects of this paper, M13 and M80. These
extreme blue tail (EBT) clusters form an important subclass of BT
clusters. All clusters found to date with EBTs have indistinguishable
\feh, close to $-1.5$. However, not all BT clusters with this
metallicity have EBTs. For example, M79  has a BT
which extends 3\,mag below the level of the HB but still falls far
short of the He-burning main sequence (\cite{hillm79}).  The
bluest stars in EBTs are often separated
from their slightly cooler counterparts by a gap analogous to that seen for
the subdwarf B stars in the Galactic field.

The distribution in color along the HB is understood to be due to a
distribution in the envelope masses of stars leaving the red-giant
branch (RGB; see Rood 1973, Buonanno et al. 1985, \cite{fpBT}).
Clusters with extended blue HB sequences often show gaps whose
physical origin is still a mystery.  They cannot be produced by
standard single population HB simulations except as statistical
artefacts (\cite{cro88}, \cite{cat97}, but see also the claims by
\cite{ldz94}). If gaps are observed at the same location in more than
one cluster HB sequence, then statistical fluctuation is unlikely to
be the cause. The reality and origin of gaps has been the subject of
several studies (\cite{rc85}, \cite{bcf85}, \cite{cro88},
\cite{bailyn92}, \cite{cat97}).

It is important to bear in mind, though, that even though gaps in
cluster HBs may be most apparent in the clusters at metallicity \feh
$= -1.5,$ but their causes may still be active in clusters at
different abundances.  In metal-rich clusters, the HBs are collapsed
to the red, and gaps that might be present are obscured. In the most
metal-poor regime, the sensitivity of a star's initial position on the
ZAHB to envelope mass ($d\log\,T_{\rm eff}/dM_{\rm env}$) is reduced.
A gap in mass $\delta M_{\rm env}$ that produces a gap at $\feh =
-1.5$ would not necessarily produce a gap at $\feh = -2.3$. Instead, a
dip in the stellar density might result as suggested for M92 by
Crocker \etal\ (1988).

 Apart from age variations (see the discussion in the review by
Stetson \etal\ 1996), theoretical studies of HB morphology all require
assumptions about the amount and distribution of mass loss on the RGB.
Unfortunately, the precise understanding of mass loss from cool stars
is one of the most vexing problems in stellar evolution theory.
Indeed, the basic underlying mechanism(s) for mass loss has (have) not
been firmly identified. Here, we adopt as a working hypothesis that
{\it there is a relationship between gaps and mass loss.} This
suggests that either there is more than one mass loss mechanism or
that a single mechanism can lead to a multiplicity of results. By adopting
this hypothesis, we hope to advance our understanding of both
phenomena.

There are some indications that the structural parameters of the
parent cluster (see \cite{fpBT,fp96,buo97}) or stellar rotation
(\cite{r77,fpr78,prc95}) might play a role in HB morphology.  Both
could plausibly affect mass loss directly.  In addition, both might
lead to mixing phenomena occurring on the RGB (see \cite{sweigart97}).
Multimodal HBs could easily arise: additional mass loss occurs if the
rotation rate exceeds some critical value; a star visits the cluster
center or not while on the upper RGB; a star mixes or not. Because
these and other candidate processes are difficult or impossible to
model from first principles, a first crucial step in understanding is
to secure data for and make comparisons between clusters. When do gaps
occur? What are the physical parameters (\teff, \logg) of the gaps?
Comparisons must be made to explore the different dimensions of
parameter space, e.g., central cluster density at fixed abundance;
abundance at fixed density.

 In this paper, we consider the HB morphology of three clusters with
HB tails of similar metallicity ($\feh \approx -1.5$) but different HB
tails.  M13 and M80, using data taken from our Cycle 5 HST/WFPC2
program (GO-5903), and M3 from our Cycle 4 HST/WFPC2 program
(GO-5420). We study the detailed morphology of the HB sequence in
far-UV/$U$/optical colors, since in UV CMDs the HB is approximately
horizontal even at high \teff.  To the extent possible, we make 
comparisons with gap locations previously observed in other clusters
(\ngc{2808}, \ngc{6752}, \ngc{6681}, $\omega$~Cen).

\section{Results}

Exposures for M80 and M13 were taken through the $U, ~B,~V$ ($=$
F336W, F439W, F555W) and the mid- and far- UV filters (F255W, F160BW)
mapping the cluster cores.  Here we use only $U,~V$, far-UV
exposures.  The CMDs presented here are {\it preliminary results} of
the four WFPC2 fields obtained with the the PC located on the cluster
center.  The detailed description of the exposures and data reduction
procedure will be given elsewhere (see \cite{m3bss} for a brief
description).  All the instrumental magnitudes have been converted to
a fixed aperture and then calibrated to the STMAG system using Table 9
in Holtzman et al (1995).

\begin{figure}[htb]
\plotone{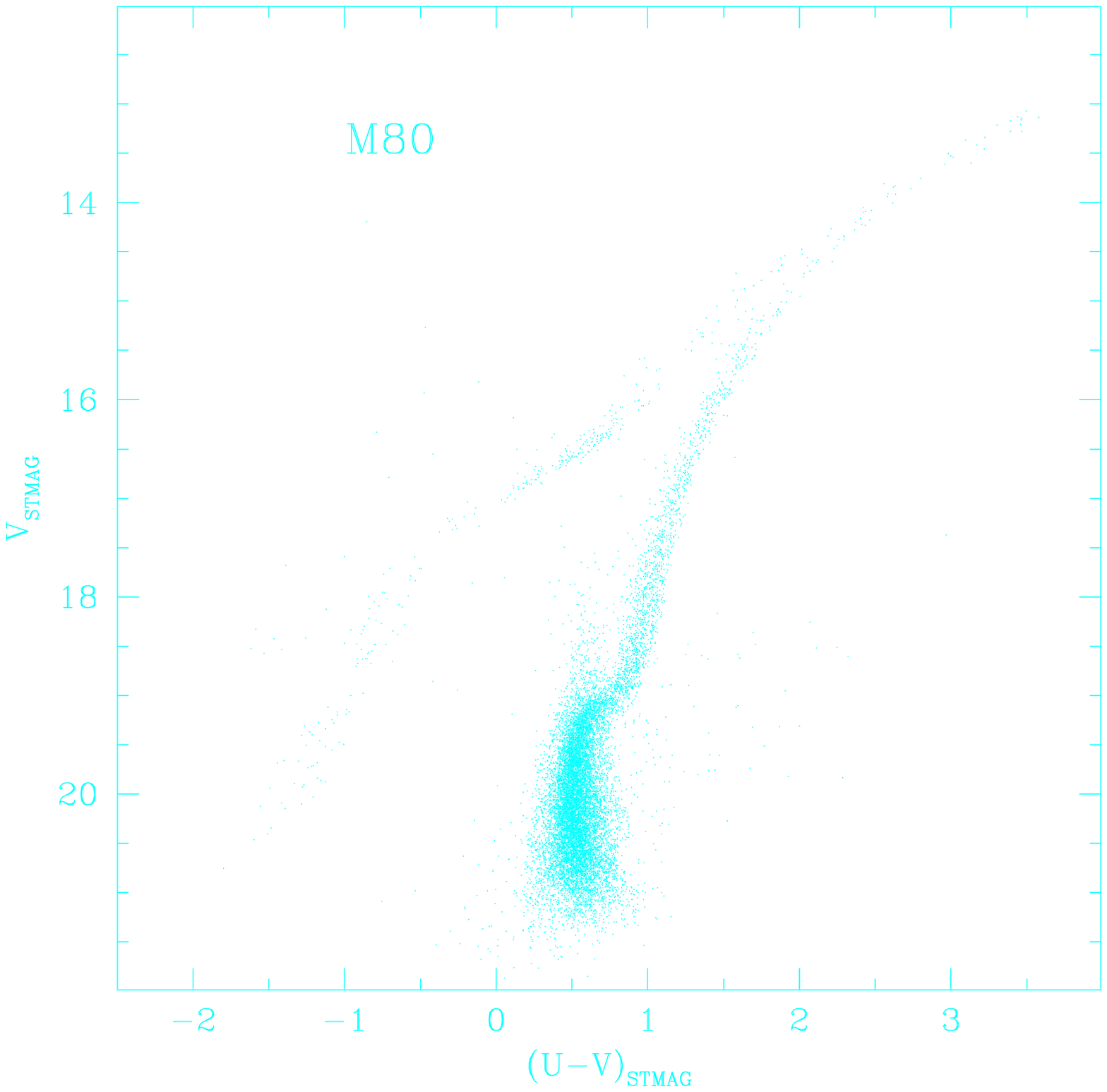}
\caption{\small
\label{m80}
\textit{The $(V,~U-V)$ color magnitude diagram for NGC6093 $=$ M80.}
}
\end{figure}

Figure~\ref{m80} shows the $(V,~U-V)_{\rm STMAG}\footnote{All
magnitudes quoted here are in the STMAG system, i.e. $m_\lambda =
-2.5\log F_\lambda - 21.1,$ where $F_\lambda$ is the flux per unit
wavelength received through a filter with effective wavelength
$\lambda.$}$ color-magnitude diagram (CMD) for more than 13,000 stars
identified in the central region of M80. All the stars found in the
PC and WFs cameras are plotted. The reader is referred to Figure~1 of
Ferraro \etal\ (1997b) for the CMDs of M13 and M3.

The most notable features of the M80 CMD are: (1) a very long blue
tail of the HB, extending $\sim 4.5$ mag (just as seen in M13,
\ngc{6752} and the blue side of \ngc{2808}); (2) a non-uniform
of the stellar distribution along the HB, with at least 4 groups of stars
separated by gaps at $V \sim 17.1$, 17.5, and 18.8.
(3) a relatively large population of supra-HB stars, which will be discussed
further in Dorman \etal\ (1997b); and (4) a large population of blue
stragglers, more similar to that found in M3 than to M13
(\cite{m3m13}).  We anticipated a substantial UV bright population on
the basis of integrated UV colors (\cite{vdd81,dB85}).  The long blue
tail of the M80 HB has been previously observed in the deep $(U,~U-B)$
CMD of Shara and Drissen (1995). Their HB extension (ranging from $U\sim
17$ down to $U\sim 19$) agrees well with that found from our CMD.
However, the four groups of HB stars clearly visible in our Figure 1
can be barely identified in the Shara \& Drissen CMD, perhaps due to their
use of aperture photometry which typically gives larger errors than
PSF fitting.

Recently, the long blue tail of the HB has also been observed in a 
ground-based study of the outer parts of the cluster (\cite{broc97}), but 
again the gaps are only barely visible in that diagram.

\section{The Detailed HB Structure of M3, M13 \& M80}

The metallicity of M80 is similar to that of M3 and M13. The Zinn
(1985)  scale gives the metallicities of  M3, M13, and M80 respectively as
(1986)  $\feh=-1.66$, $-1.64$, and $-1.65.$ Note that new high resolution
spectra suggest higher \feh\ values for M3 and M13, $\feh
\sim -1.47$ and $-1.51$, (Kraft \etal\ 1992) and $\feh \sim -1.34$ and
$-1.39$ (Carretta and Gratton 1997 [CG97]). We assume that the 
determination of the abundance differential between the clusters is more robust than
the absolute value. Using equation 7 in CG97 to convert the Zinn value to CG97 scale
gives $\feh\sim -1.4$ for M80.

To get a better impression of the similarities in the overall morphology in
the CMD we align the clusters in the $(V,~U-V)_{\rm STMAG}$ CMD,
shifting the CMDs to match the M13 principle sequences and then
co-adding the result. The shifts necessary are: 
$\delta V=-1.3$ and $\delta (U-V)\sim -0.3$ for M80 and 
$\delta V=-0.6$ and $\delta (U-V)\sim -0.03$  for M3. 
The color shifts imply $\Delta{E(B-V)}_{M80-M13} \lesssim 0.15$
 and $\Delta{E(B-V)}_{M3-M13} \lesssim 0.015$ (see Dorman \etal\ 1997b for further discussion).

\begin{figure}[htb]
\plotone{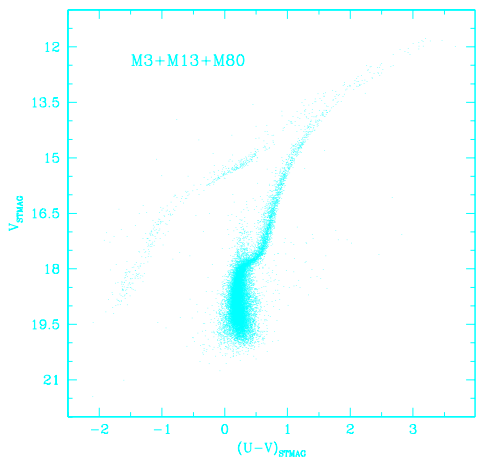} 
\caption{\small
\label{cmdall}
\textit{The coadded 
$(V,~U-V)$ 
CMD for  M13, M80 and M3. M80 and M3 have been aligned with M13 shifts
in $(U-V), V.$ See text for details.}
}
\end{figure}

Figure~\ref{cmdall} 
shows the combined CMD, in which more than 50,000 stars have
been plotted. As can be seen from the small scatter along the RGB and
HB the sequences match well, showing the high degree of similarity of
the main branches in the CMD for these clusters. Moreover the HB
multi-group distribution is still (perhaps more obviously) present.
For this to be the case the gaps along the HB of these clusters must
have very similar locations in the CMD.  In addition, the HB of M3
nicely matches the reddest part of the HB in M13 and M80. Note that in
this plane the RR Lyrae gap occurs at $(U-V)\sim 0.55$ and the red
HB covers a quite small range in color, since the $(U-V)$ color is
quite insensitive to $\teff$ for temperatures less than or equal to
those of RR Lyrae stars (\cite{m3m13}).

\begin{figure}[htb]
\plotone{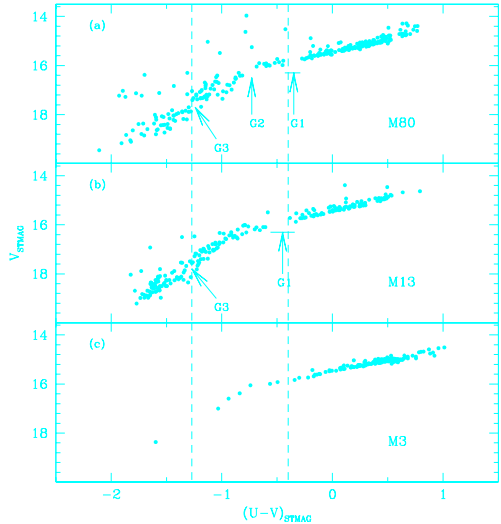}
\caption{\small
\label{hbuv}
\textit{The HB in the $(U-V,~V)$ plane, for M3, M13 and M80, respectively,
after the alignment to M13 (see text). Only non-variable stars
have been plotted for M3. 
Gaps discussed in the text are shown.
Note that in this plane the RR Lyrae
gap occurs at $(U-V)=0.55$ and it is not visible in the CMD
since it spans a small range in color.}
}
\end{figure}

In order to bring out the similarities, we show in Figure 3 a
comparison of the HB of each cluster in the $(V,~U-V)$ plane after
alignment. The relative population of HB stars in the three clusters is
292, 230 and 163 in M80, M13 and M3, respectively (only non-variable
stars have been considered). These numbers reflect the
larger mass estimated for M80 and the large fraction of M3 stars which
is variable.

\begin{figure}[htb]
\plotone{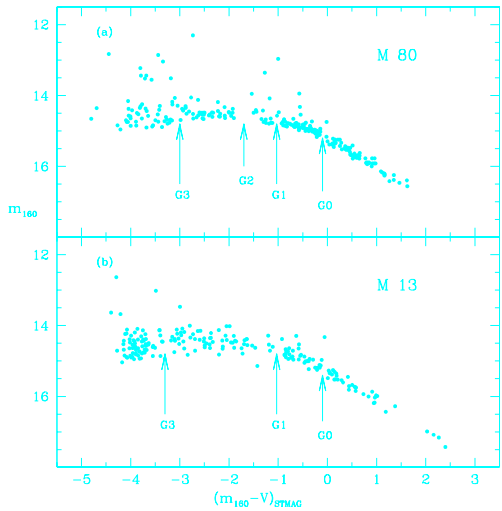}
\caption{\small
\label{160cmd}
\textit{The HB in the $(m_{160},~m_{160}-V)$ plane, for M13 and M80, respectively.
A shift of
$\delta m_{160}=-1.95$
and $\delta (m_{160}-V)\sim -0.75$ 
has been applied to M80 HB to match M13.
Gaps discussed in the text are shown.}
}\end{figure}

\begin{figure}[htb]
\plotone{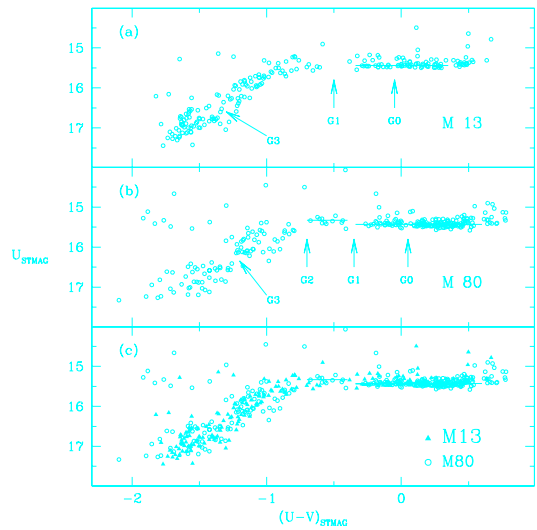}
\caption{\small
\label{uuvcmd}
\textit{The HB in the $(U,U-V)$ plane, for M13 [{\it panel (a)}] and M80
[shifted to match M13 {\it panel (b)}], respectively.
The luminosity of the stars between G1 and G2 (with respect to the
``horizontal'' region of the HB in this plane) is also shown.
In {\it panel (c)} the HB stars in M80 (empty circles) from {\it panel (a)}
have been plotted together with the HB stars in M13 (filled triangles), 
for comparison.}
}
\end{figure}

Figure~\ref{160cmd} shows the comparison of the HB of M13 and M80 in the
($m_{160}$, $m_{160}-V$) plane. A shift of $\delta m_{160}=-1.95$ and
$\delta (m_{160}-V)\sim -0.75$ has been applied to M80 to match the
M13 sequence. Figure~\ref{uuvcmd} shows the $U,~U-V$ CMD.

The gaps are more obvious in some planes than in others, but each gap we
discuss is present in each plane. Several important
features can be seen in these figures. First, the location of the cooler
gap in M13 and M80, labeled G1, is almost coincident with the blue edge of
the bulk of the HB population in M3.  Second, the hottest gap is found in
similar positions in both M13 and M80.  Third, M80 appears to have
 another gap, G2. In Figure~\ref{uuvcmd} we see that the M80 stars
between G1 and G2 define a sequence somewhat more luminous than that of the
adjacent regions or in the comparable part of M13 (see panel c). Finally, 
Figures~\ref{160cmd} and \ref{uuvcmd} show an additional gap, G0, 
located at $(m_{160}-V) \sim 0$. G0 is less apparent in the
$V,~U-V$ CMD because of the compression of the
\teff\ scale in the $U-V$ color.  However, careful examination of Figure~3
does reveal features which can be associated with G0. 

A preliminary
comparison with the theoretical models (Dorman \etal\ 1993) shows that G1 is
located at $\teff \sim 11,000\kelvin$, G2 at $\teff \sim 12,000\kelvin$
and G3 at $\teff \sim 19,500\kelvin$ in M13 and $\teff \sim 18,000\kelvin$
in M80, and G0 at $\teff \sim 9,500\kelvin$. It is quite remarkable that
G0, G1, and G3 are located at nearly the same temperature (within
1500\kelvin) in each cluster.
 
\section{ \lhb: A Linear Coordinate along the HB}

In order to study the stellar distribution along the HB properly 
and to test for the existence and significance of features like gaps, Rood
\& Crocker (1985, hereafter RC85) and Crocker \etal\ (1988) introduced
a coordinate which was ``linear'' along the entire length of
the HB.  Such a coordinate was required to avoid
``artificial'' features (clumps, etc.) generated by the saturation of
the color index as it becomes insensitive to changes in the
temperature.

RC85 defined a $X_{\rm HB}$ coordinate which measured the star
position along the HB as its projection on the theoretical ZAHB (see
Figure~1 in RC85).  Analogously Ferraro \etal\ (1992a,b), Dixon \etal\
(1996), and, more recently Catelan \etal\ (1997) have used a coordinate
$\lhb$, based only on observable quantities. \lhb\ utilizes the mean
ridge line for the linear coordinate along
 the HB in the observational plane. We will use
\lhb\ to study the peculiar distribution along the
HB tail for the M13, M80 pair in more detail.

Because of the high degree of similarity of the main branches in M13 and
M80 (see Fig 2), as first step we have selected the non-evolved HB
stars in the $(V,~U-V)$ plane in the co-aligned sample.
A total sample of 492 stars (221 in M13 and 271 in M80) has been fit
by a 6th order polynomial to define the mean ridge line. 
Each star was then projected onto the ridge line, and a curvilinear coordinate
($\lhb$) has been computed along it. The zero point of the $\lhb$
coordinate has been arbitrarily taken to be $(U-V)=1$.

\begin{figure}[htb]
\plotone{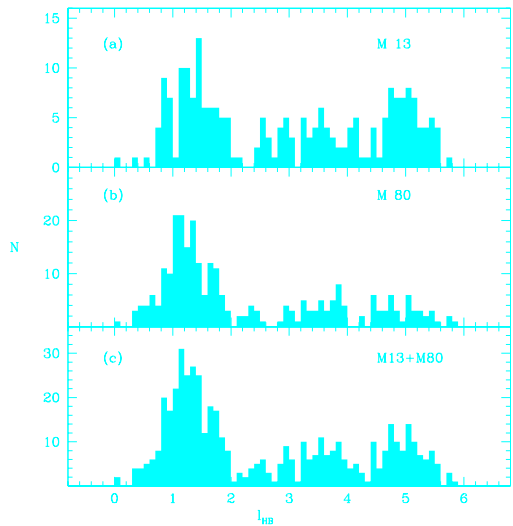}
\caption{\small
\label{lhb}
\textit{Histograms showing the $\lhb$ distribution for M13 [{\it panel (a)\/}],
M80 [{\it panel (b)\/}] and
for the co-added sample $({\rm M13+M80})$; [{\it panel (c)\/}].
The zero point of the $\lhb$ coordinate has been assumed at 
$(U-V)=1$, so low values of $\lhb$ correspond to the red side of the HB,
high values to the extreme blue tail.}
}
\end{figure}

The \lhb\ distributions for each cluster and the co-added sample are
shown in Figure~\ref{lhb}.  From the inspection of panels (a) and
(b) in Figure~\ref{lhb}, it is evident that the stellar
distribution along the tail of these clusters is quite complex. Despite the fact that
neither HB is uniformly populated and the HB distribution is quite different
in the two clusters, there are some striking similarities in their overall morphology.
These similarities are generally best seen in the co-added sample displayed in
panel (c). Note in particular:

\begin{enumerate}

\item The global length of the two HBs is approximately the same 
($\lhb \sim 6$); only one star in M80 (located at $U-V=-1.8$) has been
found beyond this limit.

\item The bulk of the reddest stars ends at the same location $\lhb \sim 2$.

\item The bluest groups start at the same location $\lhb \sim4.4$.

\item 
The co-added sample shows at least three main groups of stars that could be
modeled individually by separate gaussian distributions with different means
and dispersions. These groups are separated by two underpopulated regions (at
$\lhb \sim 2.1$ and $\lhb \sim 4.3$). Additional ``depressions'' in the star
 counts are located at $\lhb \sim 2.7$ and $\lhb \sim 3.2$.

\end{enumerate}

In the following we denote the three main groups of stars as follows:

\begin{description}

\item[UBT] Upper blue tail HB stars: $\lhb <2.1$

\item[MBT] Mid-blue tail HB stars: $2.1< \lhb <4.2$

\item[LBT] Lower blue tail HB stars $\lhb >4.2$. It is the presence of
a substantial population of LBT stars which sets EBT clusters apart
from other BT clusters.

\end{description}

\noindent
The relative population of the HB
stars in these groups, and the values are given in Table~1.
Interestingly, the fraction of stars in the LBT is
much larger in M13 ($\sim 30\%$) than in M80 ($\sim 17\%$).

\begin{deluxetable}{lccl}
\tablewidth{\textwidth}
\label{btdist}
\tablecaption{Relative populations of stars along the HB (see text)}
\tablehead{
 \colhead{}           & \colhead{UBT}      &
\colhead{MBT} & \colhead{LBT}
}
\startdata
 M13 & 92 & 62 &  67 \nl
 M80 & 169 & 57 &  45 \nl
\enddata
\end{deluxetable}

\subsection{The gaps}

Catelan \etal\ (1997) have recently discussed the problem of determining
the statistical significance of the gaps.  From an extensive set of the
synthetic HBs they found that the procedure described by Hawarden (1971)
and Newell (1973), often adopted to evaluate the significance of
gaps (e.g., Ferraro \& Paresce 1993, Crocker \etal\ 1988), substantially
overestimates their statistical significance.  As a first attempt to
quantify the probability of finding gaps along the blue HB, they
assumed that the stellar distribution in $\lhb$ coordinate is uniform, and computed 
the probability of a gap of given depth (number of stars in a given range of $\lhb$ less
than some value) anywhere along the HB sequence. Under this idealized assumption the probability
is naturally governed by the Binomial distribution

\begin{figure}[htb]
\plotone{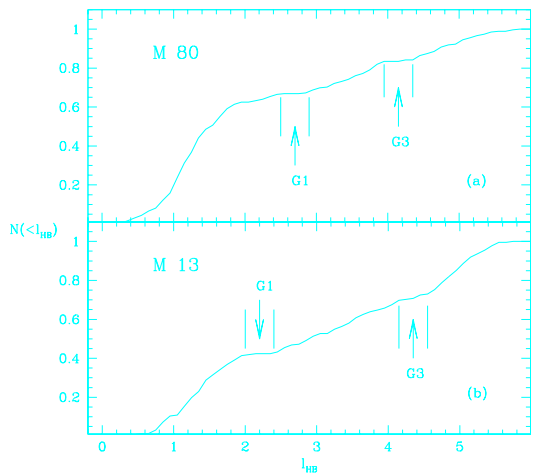}
\caption{\small
\label{lhbcum}
\textit{The $\lhb$ cumulative distribution of the HB stars in 
M13 [{\it panel (a)\/}] and 
M80 [{\it panel (b)\/}]. Gap locations gaps are indicated by arrows.}
}
\end{figure}

To follow their procedure we first plot in Figure~\ref{lhbcum} the
normalized cumulative \lhb\ distribution for M13 and M80.
In this plot the gaps should appear as horizontal regions, and their
location is indicated by arrows.  The typical size of the gaps in this
coordinate is $\delta \lhb \sim 0.4,$ and the HB sequences has length
up to $\lhb \sim 6.$ Divide the BT into 15 bins of width  $\delta \lhb \sim 0.4.$
Given the total number of HB stars and the number of stars actually found
 in each gap, one can evaluate the probability that a gap of similar depth would
be found somewhere along the blue HB.  The bluest gap in M13 (G3) has quite a low statistical
significance, with the probability that it is due to chance $\sim20\%$. However, the probability that 
this depression in the star counts occurs {\it close to the same $\lhb$ value in both clusters} remains
small.  All the other gaps that can be seen in the $(V,~U-V)$ CMD, are deep enough to have
quite high a statistical significance ($>3.5 \sigma$).

\section{Discussion}

The variation of HB morphology among clusters of similar metallicity
is known as the {\it second parameter} (\SP) problem. However, all of the
complexities of HB morphology may not be so easily classified. 
The \SP\ problem has aspects that require comment here as a result of these 
new observations.

\begin{enumerate}

 \item There is evidence in favor of a ``global'' \SP\
(\cite{ldz94}, \cite{fpBT}). By ``global'' we mean some process or
circumstance that acts to vary HB morphology at fixed metallicity in
the cluster system as a whole. We exclude from ``global'' 
individual differences between cluster HB morphology that might
arise through exceptional circumstances in certain clusters (e.g.,
\ngc{2808}). One possible global \SP\ that has been explored extensively is age
(\cite{ldz94}, \cite{cds96} and references therein), which shifts the
peak of the HB color distribution to the blue.

 \item Parameters that characterize the bulk HB morphology
(\cite{fpBT}) do not describe the overall extension of the BT.  BTs
may be driven by a special ``\SP,'' which we will call the \btsp. Fusi
Pecci \etal\ (1993) and Buonanno \etal\ (1997) noted a strong
correlation between the blue tail extension and the cluster central
density, which they suggested as ``a'' \SP.  Stellar rotation
(\cite{prc95}) and helium mixing (\cite{sweigart97}) are other
plausible \btsp s.  In addition, there is a subset of the BT clusters
with highly extended blue tails (EBT clusters) in which mass loss goes
to completion in a significant fraction of the HB stars. Until
recently, \ngc{6752} was the only known example of this morphology
(\cite{cannon81}), but in the last few years it has been joined by
$\omega$ Cen (\cite{wh94}), \ngc{2808} (\cite{sos97}), in addition to
M13 and M80.

 \item Features like the BT gaps may be regarded as \SP\ ``fine
structure.''  Our tentative working assumption is that this fine
structure is imposed on the HB by RGB mass loss.  Mass loss
``structure'' could well be linked to different physical processes
like cluster core interactions or stellar rotation, or perhaps by
other unknown factors---i.e., whatever is driving the BTs is also
establishing the HB fine structure. The results of the processes
causing these fine structure phenomena may look very different in
clusters of higher and lower metallicity.

\end{enumerate}

In Ferraro \etal\ (1997b), we made a comparison of the HB morphology
of M3 and M13---an extreme example of a second parameter pair.  Since
the relative locations of the main loci (RGB, HB, MS-TO) of those
clusters was quite similar, we argued that that {\it age cannot be
responsible for the HB differences observed in M3 and M13}.
Since the fiducial sequences of M13 and M80 can also be aligned,
there is {\it no reason to infer that the age of M80 is significantly different
from that of M3/M13}.

We can now bring M80 into the discussion to illuminate further the possible
role of cluster dynamics on HB morphology.  M3 and M13 have almost the
same central density ($\log \rho_0 \sim 3.4 \msun\,{\rm pc}^{-3}$) but
different HB morphology (Ferraro \etal\ 1997b). M80 is one of the most
concentrated clusters in the Galaxy ($\log \rho_0 \sim 5.3
\msun\,{\rm pc}^{-3}$) which has no evidence for a core collapsed state
(i.e. its surface brightness profile can be modeled by a King-Michie model).
Despite the fact that M80's core density is almost two orders of
magnitude higher than that of M13, its HB morphology is very similar.
Thus, either cluster density is not affecting HB structure in
M13 \& M80, or different \btsp s are at work in the clusters.
Coupled with the fact that density cannot account for the M3/M13
difference, this appears to rule out density as the sole \btsp.

Nevertheless, the correlation between BTs and high density seems very
strong and cannot be discounted. Almost every high density cluster
with $\feh \sim -1.5$ has a blue tail.  Even some high metallicity,
high density clusters (\ngc{6388} and \ngc{6441} \cite{rich97}) have
now been found to have the analog to a BT. In \ngc{6388}, for example,
$\sim 16-20\%$ of the stars in core have suffered a large degree of
mass loss.  Yet 47~Tuc (\cite{uit47}), which is almost as dense, has
only a small number of blue HB stars, very few of which are in the
core.

Beyond the observed correlation, an attraction of high cluster density
as a \btsp\ is that it is easy to imagine that a multi-modal
distribution blue stars might be produced through tidal interactions.
One possibility was presented by Sosin \etal\ (1997) discussing the
case of \ngc{2808}.  They argue that the EBT with gaps in that cluster
might arise through groups of stars undergoing events that strip
discrete amounts of mass from RGB envelopes. The expectation would be
that the most extreme mass loss happens to the stars with the highest
number of interactions. A consequence of this mechanism is that the
bluest stars would be relatively rare, which is not what is observed.

Likewise high density should be correlated with the number of capture-binaries
(\cite{bailyn92}).  While binary evolution might account for
sparsely populated BTs like the few stars in M3, or perhaps the few
EHB stars beyond the bulk of the BT in M15, the substantial
BTs we discuss here are almost certainly some variant of single HB
stars. If they were not, neither population ratios (\cite{bcf85}) nor
radial distributions (Ferraro \etal\ 1997b, Whitney \etal\ 1997) would
be as observed.  So while high density could plausibly produce BTs, no
mechanism which fits the observations has yet been suggested.

Does the correlation of high density with BTs coupled with the fact
that some low density clusters have BTs require that there are
multiple \btsp s? A single mechanism might suffice if high cluster
density could produce BTs through some intermediary process.  Let us
suppose that \process causes BTs. If high density cores
could lead to \process, perhaps via tidal interactions in the cluster core,
this would explain the correlation between density and BTs. If
\process\ could also arise for other reasons, perhaps as a cluster
initial condition, this would explain low density clusters with BTs.

As for \SP\ fine structure, we have argued that most
of the gaps are statistically significant. Further, we have
found that two (G1 and G3) or maybe even three (G0, G1 and G3) gaps
are located at nearly the same temperature (within 1500\kelvin) in M13
and M80. This evidence strongly suggests that some genuine physical
mechanism produces the gaps. If a second parameter candidate could be
shown to produce such features, it would be an important step toward
resolution of this long-standing problem.

\subsection{Gaps in other Clusters}

Gaps in other BTHB clusters have been known for some time. How do
these gaps compare to ours?  Catelan \etal\ (1997) list 14 GCs that
show gaps along the HB. Here we limit our discussion to BT gaps,
excluding clusters which only show mid-HB ($=$ near the RR~Lyrae
strip) gaps. Even in many BT clusters no explicit determination of the
temperature at which each gap occurs has yet been made.  Because of
the difficulty of determining \teff\ from $B,~V$ photometry alone, we
will further restrict our discussion to clusters with available large
sample UV photometry.  These include

\begin{enumerate}

\item M79 (Hill \etal\ 1996)---Based on their UIT1
$m_{152},~m_{152}-m_{249}$ CMD we find a gap at 9900\kelvin. A gap is seen
at the same temperature in the ground based $V,~B-V$ CMD of Ferraro \etal\
(1992a), There are only a few stars hotter than 20,000\kelvin. 

\item \ngc6681 (Watson \etal\ 1994)---Catelan \etal\ (1997) also note that
the $m_{160}, m_{160}-V$ \hst\ based CMD shows a gap at 8,700\kelvin\ (see
Figure~2 in Watson \etal\ 1994), which is nearly at the same temperature as G0.
Moreover the clump of HB stars seems to have a break at 11,000\kelvin,
corresponding to the temperature of G1.  There are only a few stars
hotter than 18,000\kelvin\ and the hottest star is at 22,000\kelvin.

\item \ngc6752 (Landsman \etal\ 1996)---In a $m_{162}, m_{162}-V$ CMD
based on UIT2 data there is a gap at 18,000\kelvin.

\item M15 (Ferraro and Paresce 1993)---Based on ultraviolet HST-FOC
observations, there is a gap at $\sim 9000 $K (again nearly at the same
temperature as G0). The HB terminates at $\sim 15,500$\kelvin.
 Metallicity is lower: $\feh = -2.15$ (\cite{z85})

\item \ngc2808 (\cite{sos97})---There are gaps at
25,000\kelvin\ and 17,000\kelvin\ (recall that the hottest gaps in M13
and M80 are at 18,000\kelvin and 19,500\kelvin\ respectively). The
\cite{z85} metallicity is somewhat higher: $\feh = -1.37,$ but similar and
within the uncertainties of abundance determinations.

\item $\omega$~Cen (Whitney \etal\ 1994)---In a $m_{162}, m_{162}-u$ CMD
based on UIT1 data there is a gap at 16,000\kelvin. Interpretation of
$\omega$~Cen is complicated because of the spread in metallicity among the
stars and because of the unexplained
depression in UV luminosity of the hottest stars in the cluster (Whitney \etal\ 1997). 
The presence of a gap does suggest that whatever mechanism produces
the $\omega$~Cen gap may also produces a gap at similar \teff\ independent of
metallicity. 

\end{enumerate}

\noindent Except as noted metallicities are all close to $\feh = -1.5$.
These results are summarized in Table~2. A ``Not EBT'' entry denotes that
the extended part of the BT is not substantially populated.

\begin{deluxetable}{lcccl}
\tablewidth{\textwidth}
\label{gaptab}
\tablecaption{Temperature of gaps in different clusters}
\tablehead{
 \colhead{}           & \colhead{G0}   &
\colhead{G1} & \colhead{G2}
& \colhead{G3} 
}
\startdata
 M13 & 9500K & 11,000K &          & 19,000K \nl
 M80 & 9500K & 11,000K & 12,000K  & 18,000K \nl
 NGC 6681 & 8700K  & 11,000K (?)  &  & Not EBT \nl
 M79 & 9900K  &   &  & Not EBT \nl
 NGC 6752 &  &   &  & 18,000K \nl
 M15 & 9000K  &   &  & Not EBT \nl
 NGC 2808 &  &  & 17,000K  & 25,000K \nl
 $\omega$ Cen &  &   &  & 16,000K \nl
\enddata
\end{deluxetable}

Identifying trends in such a compilation is difficult because: 1)
There is ``noise'' in the gap data. Catelan \etal\ argue that gaps at
random locations along the BT are not terribly unlikely. Hence we may
be confused by a few statistical fluctuation gaps mixed in among
``real'' gaps. For example, G2 in M80 (but see below) and G3 in
\ngc{2808} might fall into this class since they have no counterparts
in other clusters. 2) The data for different clusters is not very
homogeneous.  3) Gaps cannot be located with high precision.
Nevertheless, some conclusions that emerge are:

\begin{enumerate}

 \item All clusters with EBTs have a at least one gap on the lower BT. 

 \item If one allows for some imprecision in the $\omega$~Cen results
there is a suggestion that all EBT clusters have a gap (G3 in M13,
M80, \ngc{6752}, and G2 in \ngc{2808}) at $\sim 18,000$\kelvin. We
will call this the 18kK gap.

 \item G0 \& G1 occur in many but not all clusters. The location and
widths of these gaps may vary cluster to cluster, but are at similar
locations in the clusters in which they are present.

\end{enumerate}

\subsection{Blue Tails and Gaps: A Naive Working Scenario}

HB/post-HB evolutionary tracks change morphology at a \teff\ similar
to that of G3. The hotter stars (so-called extreme-HB or EHB) evolve
essentially vertically in the H-R diagram and do not return to the
asymptotic giant branch (AGB) after core He exhaustion. The entire
post-HB evolution is spent at high \teff\ as \manque\ stars
(\cite{dor93,gr90}). The cooler stars return to the AGB.  It is
tempting to identify the ``18kK'' gap with the onset of EHB behavior
(Newell 1973).  Rood \etal\ (1997a,b) show that the transition from
normal HB track morphology to EHB morphology does not produce a gap.
However, a combination of this factor plus considering mass loss to be
function of a random (but unimodal) mass loss efficiency (\etaml) as
suggested by D'Cruz \etal\ (1996) can produce a gap (Whitney \etal\
1997).  Basically, a large range in \etaml\ produces EHB stars and
thus a pile-up on the hot end of the HB. Coupled with a sparse
population along the rest of the BT, this can lead to gap near but not
necessarily at the normal-HB-EHB transition. While the D'Cruz \etal\
(1996) result was based on a specific model, they argued that the
result was more general and that the distribution of stars below the
EHB gap could lead to observational evidence as to how the RGB mass
loss turns-off (Rood
\etal\ 1997a).

The gap we have labeled G0 occurs frequently enough that it may be
real. It occurs near the hot end of the traditional ``horizontal'' (in
the $V,~B-V$ CMD) HB.\footnote{Just as many white dwarfs are not
white, we now know that many HBs are not horizontal. In this context
we are drawn to the usage horizontal horizontal branch.} This gap is
present in some but not all clusters.

The gaps G1 and G2 are more complicated because G2 appears in M80 but not
M13. We suggest the following scenario (Refer to all three panels
of Figure~\ref{uuvcmd} for the following discussion.) Our working
hypothesis is that BT-gaps arise because RGB mass loss does not uniformly
populate the ZAHB. It seems unlikely that a mass loss process would
produce a ZAHB region completely devoid of stars, and more likely that there would be
a mass loss range with low probability. This would produce a sparsely
populated region of the ZAHB, which in combination with small number
statistics could produce a gap. We suggest that the entire G1--G2 region
is a ``sparse population gap'' (recall that the M80 stars in this region are 
slightly overluminous with respect to the cooler stars). 
Such a gap might contain a few near ZAHB
stars. In addition, it might contain stars which originated from the
more richly populated regions on the hot side of the gap that are now
evolving back toward the AGB. The latter are analogous to the RR~Lyrae
population in clusters like M92 and would be more luminous than the bulk
of the HB stars. (This group would not be present in G3 assuming our
identification of G3 as the EHB gap is correct.) As luck would have it most
of the G1--G2 stars in M13 are from the sparse ZAHB group, and those in M80
have evolved from the hotter parts of the HB.

No explanation for the G0 and G1/G2 leaps forward. One possibility is
that there are two mass loss mechanisms, ML1 and ML2, driven by
different engines.  In all clusters mechanism 1 operates and produces
a mean mass loss of $\sim 0.2$\msun. ML1 almost certainly is a
function of metallicity.  For example, Fusi Pecci \etal\ (1993) argue
that very metal-poor stars probably lose less total mass.  In clusters
where ML1 is the predominant mass loss mechanism, one observes normal
\FP\ behavior. In addition, a global \SP\ like age would produce the
expected behavior, e.g., increasing age $\Rightarrow$ bluer HB.

A second mechanism operates to different degrees in different
clusters and on individual stars within each cluster. In some clusters
it is insignificant compared to ML1. Where it is significant, ML2 is
basically the \btsp.  ML2 is probably driven by different engines
including cluster density and stellar rotation. It may be accompanied
by subsidiary effects like He-mixing which lead to diverse outcomes.
If ML2 operates very efficiently, some HB stars are effectively removed 
from the distribution of ML1-only HB stars producing a gap. If ML2 is only
mild, the ML1 + ML2 stars might form a tail merging into the ML1-only
HB. This might fit the bimodality-without-a-gap suggested by Crocker
\etal\ (1988) for clusters like M92. 

Whether an alternate mechanism could mimic ML2 without additional mass
is not clear. In the Sweigart (1997) He-mixing models, substantial
mixing was usually accompanied by substantial mass loss, although the
free-parameter space is large enough that this might not be required.
In general though, it is easier to modify the composition of a RGB
envelope if most of the envelope has been lost.

Clearly, building a case for a scenario like that above will require data
from more clusters and a more precise determination of the gap parameters
in each cluster. Also requiring further study is the nature of the gaps.
Are there forbidden zones which should be completely devoid of stars
except those scattered in by observational error?  Perhaps as suggested
above, they are just underpopulated regions? 

\subsection{Other Related Factors}

There are other cluster phenomena which may be related to HB
morphology. Even if a causal connection cannot be established it is
worth searching for correlations.

The phenomenon most likely correlated with BTs is RGB mixing as shown
through the abundances of the CNO elements. \cite{catelanm3m13} 
noted the difference in RGB CNO processing in their study of M3 \&
M13. They searched for similar correlations in other clusters. In
particular they note that the most highly processed super-oxygen-poor
stars (SOP) of Kraft \etal\ (1992) are found only in clusters with very
blue HBs. As far as we are aware, the required abundance study of M80
has not been done. We might predict on the basis of HB
morphology that M80 will have SOP stars.

\cite{norris83} noted a possible
correlation between HB-morphology and cluster ellipticity. For the
M3/M13 pair this seemed an interesting possibility (Ferraro \etal\
1997b). For M80 the ellipticity is 0.06, compared to 0.01 \& 0.12 for
M3 \& M13, respectively. So the correlation is neither strengthened
nor weakened since M80 could be inclined to our line of sight.

Another perhaps remote possibility is a correlation with the
parameters of the clusters' orbits in the Galaxy (e.g., Dauphole
\etal\ 1996). The current sample of cluster orbits is small and does
not include M80.

\section{Summary}

Using \hst\ UV/visual observations we have found that the HB of the
globular cluster M80 has a long blue tail extending below the MS-TO.
The stellar distribution along the HB is not uniform, but contains
four gaps. M80's HB is strikingly similar to that which we have found
earlier in M13. In particular,  the locations of three of the
gaps are similar. Because of this we argue that the gaps are real, i.e., they are
not statistical fluctuations, and that the gaps owe their origin to
some yet unidentified physical process---most probably associated with
RGB mass loss.

The role of central density in determining the HB morphology of M80, M3,
and M13 is not at all simple. M13 and M3 have similar (lowish) densities
and very different HBs. The density of M80 is almost two orders of
magnitude higher and its HB morphology is quite similar to that of
M13.  The correlation of density and HB-BTs in other clusters is quite
strong.  We present a tentative working scenario: first, the \SP\
problem is perhaps not a single game (e.g., cricket and
baseball---similar sports which are indistinguishable and
incomprehensible to non-aficionados); second, there many players
participating in the \SP\ games, to different degrees from
cluster-to-cluster and star-to-star.  Though this may seem to be {\it
ad hoc}, we note that nature is complex and that these stars have
lived crowded together for billions and billions of years.

\acknowledgments

RTR \& BD are supported in part by NASA
Long Term Space Astrophysics Grant NAGW-2596 and STScI/NASA Grant
GO-5903. The financial support by the {\it Agenzia Spaziale Italiana}
(ASI) is gratefully acknowledged.

\onecolumn

\end{document}